\documentclass[10pt,conference]{IEEEtran}
\usepackage[USenglish]{babel}
\usepackage{graphicx}
   \DeclareGraphicsExtensions{.eps}
\usepackage{amsmath}
\usepackage{amsfonts}
\newcommand{\fdk}{f_{d,k}}
\newcommand{\pth}{p_{\text{th}}}
\newcommand{\f}{\mathbb{F}}
\newcommand{\dmin}{d_{\text{min}}}
\newcommand{\ux}{\underline{x\!}}
\newcommand{\ualpha}{\underline{\alpha}}
\newcommand{\ua}{\underline{a}}
\IEEEoverridecommandlockouts  % to allow \thanks command in conference mode
\title{Extended Non-Binary Low-Density Parity-Check Codes over Erasure Channels\vspace{-1mm}}
\newlength{\autvspace}
\author{%
    \IEEEauthorblockN{Lam PHAM SY}                                      %
    \IEEEauthorblockA{\small EUTELSAT SA  \\                            %
                             F-75015, Paris, France \\                  %
                             lphamsy@eutelsat.fr \vspace{\autvspace}}   %
\and %
    \IEEEauthorblockN{Valentin SAVIN}                                   %
    \IEEEauthorblockA{\small CEA-LETI, MINATEC campus  \\               %
                             F-38054, Grenoble, France \\               %
                             valentin.savin@cea.fr \vspace{\autvspace}} %
\and %
    \IEEEauthorblockN{David DECLERCQ}                                   %
    \IEEEauthorblockA{\small ETIS, ENSEA / Univ. Cergy-Pontoise / CNRS UMR-8051    \\       %
                             F-95000, Cergy-Pontoise, France   \\       %
                             declercq@ensea.fr \vspace{\autvspace}}     %
\thanks{This work was supported by the French National Research Agency (ANR), grant No 2009 VERS 019 04 -- ARSSO project.}%
}

\begin{document}
\maketitle

\begin{abstract}
Based on the extended binary image of non-binary LDPC codes, we propose a method for generating extra redundant bits,
such as to decreases the coding rate of a mother code. The proposed method allows for using the same decoder,
regardless of how many extra redundant bits have been produced, which considerably increases the flexibility of the
system without significantly increasing its complexity. Extended codes are also optimized for the binary erasure
channel, by using density evolution methods. Nevertheless, the results presented in this paper can easily be
extrapolated to more general channel models. \vspace{-2mm}
\end{abstract}

% ==============================================
  \section{Introduction}\label{sec:introduction}
% ==============================================
Adaptive coding techniques are frequently employed, especially in wireless communications, in order to dynamically
adjust the coding rate to changing channel conditions. An example of adaptive coding technique consists in puncturing a
mother code. When the channel conditions are good more bits are punctured and the coding rate is increased. In poor
channel conditions all redundant bits are transmitted and the coding rate drops. However, in harsh conditions, the
receiver might not be able to successfully decode the received signal, even if all the redundant bits have been
transmitted. In such a case, the coded block can be retransmitted until the sent information is successfully decoded.
This is equivalent to additional repetition coding, which further lowers the coding rate below the mother coding rate.

However, the use of retransmission techniques might not be suitable nor possible in some situations, such as
multicast/broadcast transmissions, or whenever the return link is strictly limited or not available (such situations
are generally encountered in satellite communications). The main alternative in this case is the use of {\em erasure
codes} that operate at the transport or the application layer of the communication system: source data packets are
extended with redundant (also referred to as {\em repair}) packets that are used to recover the lost data at the
receiver. Physical (PHY) and upper layer (UL) codes are not mutually exclusive, but they are complementary to each
other. Adaptive coding schemes are also required at the upper layer, in order to dynamically adjust to variable loss
rates. Besides, codes with very small rates or even {\em rateless} \cite{luby2002lc,shokrollahi2006rc} are sometimes
used at the application layer for fountain-like content distribution applications.

In this paper we propose a coding technique that allows to produce extra redundant bits, such as to decrease the coding
rate below the mother coding rate. Extra redundant bits can be produced in an incremental way, yielding very small
coding rates, or can be optimized for a given target rate below the mother coding rate. As for puncturing, the proposed
technique allows for using the same decoder, regardless of how many extra redundant bits have been produced, which
considerably increases the flexibility of the system, without increasing its complexity.

The proposed coding scheme is based on non-binary low density parity check (NB-LDPC) codes \cite{gall-monograph} or,
more precisely, on their {\em extended binary image} \cite{Valentin10}. If $q=2^p$ denotes the size of the non-binary
alphabet, each non-binary symbol corresponds to a $p$-tuple of bits, referred to as its binary image. Extra redundant
bits, called {\em extended bits}, are generated as the XOR of some bits from the binary image of the same non-binary
coded symbol. If a certain number of extended bits are transmitted over the channel, we obtain an {\em extended code},
the coding rate of which is referred to as {\em extended (coding) rate}. In the extreme case when all the extended bits
are transmitted, the mother code is turned into a {\em very small rate} code, and can be used for fountain-like content
distribution applications \cite{Valentin10}. A similar approach to fountain codes, by using multiplicatively repeated
NB-LDPC codes, has been proposed in \cite{kasai10}. If some extended rate is targeted, we show that the extended code
can be optimized by using density evolution methods.

%Although this paper deals only with NB-LDPC codes over the binary erasure channel (BEC), the results presented here can
%be easily extrapolated to more general channel models.

The paper is organized as follows. Section \ref{sec:nbldpc_codes} gives the basic definitions and the notation related
to NB-LDPC codes. In Section \ref{sec:extended_nbldpc_codes}, we introduce the extended NB-LDPC codes and discuss their
erasure decoding. The analysis and optimization of extended NB-LDPC codes are addressed in Section
\ref{sec:analysis_optimization}. Section \ref{sec:code_design_performance} focuses on the code design and presents
simulation results, and Section \ref{sec:conclusions} concludes the paper.

% =======================================================
  \section{Non-binary LDPC Codes}\label{sec:nbldpc_codes}
% =======================================================
We consider NB-LDPC codes defined over $\f_q$ \cite{Davey-MacKey}, the finite field with $q$ elements, where $q = 2^p$
is a power of $2$ (this condition is only assumed for practical reasons). We fix once for all an isomorphism of vector
spaces:
\begin{equation}
    \label{eq:identify}
    \f_q \stackrel{\sim}{\longrightarrow} \f_2^p \nonumber
\end{equation}
Elements of $\f_q$ will also be referred to as {\em symbols}, and we say that $\ux = (x_0,\dots,x_{p-1})\in\f_2^p$ is
the {\em binary image} of the symbol $X\in{\f_q}$, if they correspond to each other by the above isomorphism. A
non-binary LDPC code over $\f_q$ is defined as the kernel of a sparse parity-check matrix $H\in
\mathbf{M}_{M,N}(\f_q)$. Alternatively, it can be represented by a bipartite (Tanner) graph \cite{Tann} containing
symbol-nodes and constraint-nodes associated respectively with the $N$ columns and $M$ rows of $H$. A symbol-node and a
constraint-node are connected by an edge if and only if the corresponding entry of $H$  is non-zero; in this case, the
edge is assumed to be {\em labeled} by the non-zero entry. As usually \cite{Rich-Urba}, we denote by $\lambda$ and
$\rho$ the left (symbol) and right (constraint) edge-perspective degree distribution polynomials. Hence,  $\lambda(x) =
\sum_d \lambda_d x^{d-1}$ and $\rho(x) = \sum_d \rho_d x^{d-1}$, where $\lambda_d$ and $\rho_d$ represent the fraction
of edges connected respectively to symbol and constraint nodes of degree-$d$. The design coding rate is defined as $r =
1 - \frac{\int_{0}^{1}\rho(x)\text{d}x}{\int_{0}^{1}\lambda(x)\text{d}x}$, and it is equal to the coding rate if and
only if the parity-check matrix is full-rank.

% =========================================================================
  \section{Extended Non-binary LDPC Codes}\label{sec:extended_nbldpc_codes}
% =========================================================================
\subsection{Extended code description}
For any integer $1 \leq k \leq q-1$, let $[k] = (k_0, \dots, k_{p-1})^{\text{T}}$ denote the column vector
corresponding to the binary decomposition of $k$; {\em i.e.} $k = \sum_{i = 0}^{p-1}k_i2^i$, with $k_i\in\{0,1\}$. Let
$X\in{\f_q}$ be a non-binary symbol, and $\ux = (x_0,\dots,x_{p-1})$ be its binary image. The $k$-th {\em extended bit}
of $X$ is by definition:
   $$\alpha_k = \ux \times [k] = \sum_{i = 0}^{p-1} k_i x_i$$
The vector $\ualpha = (\alpha_1,\dots,\alpha_{q-1})$ is called {\em extended binary image} of $X$. Note that
$\alpha_{2^i} = x_i$, for any $0\leq i \leq p-1$. An extended bit $\alpha_k$ is said to be {\em nontrivial} if $k$ is
not a power of $2$ (hence, $\alpha_k$ is a linear combination of at least two bits from the binary image $\ux$ of $X$).

Now, consider a NB-LDPC code defined over $\f_q$, with coding rate $r = \frac{K}{N}$. Let $(X_1, \dots, X_N) \in
\f_q^N$ be a non-binary codeword, $(\ux_1, \dots, \ux_N) \in \f_2^{Np}$ be its binary image, and $(\ualpha_1, \dots,
\ualpha_N) \in \f_2^{N(q-1)}$ be its extended binary image. By transmitting the extended binary image over the channel,
we obtain a code with rate $\frac{Kp}{N(q-1)} = r\frac{p}{q-1}$, which can be advantageously used for application
requiring very small coding rates \cite{Valentin10}.

We define an {\em extension of the NB-LDPC code}, as a family of matrices $\{A_1,\dots,A_N\}$, where each
$A_n\in\mathbf{M}_{p,t_n}(\f_2)$ is a binary matrix with $p$ rows and $t_n$ columns, with $t_n\geq 0$. Let $\ua_n =
\ux_n \times A_n \in \f_2^{t_n}$; hence, $\ua_n$ is constituted of $t_n$ extended bits of $X_n$ (possibly with
repetitions, if $A_n$ contains two or more identical columns). The binary vector $(\ua_1,\dots,\ua_N)$ is called {\em
extended codeword}, and the {\em extended coding rate} is given by $r_e = \frac{Kp}{T}$, where $T = \sum_{n=1}^N t_n$.

Note that the above definition is very broad, and it can yield extended rates below as well as extended rates above the
mother coding rate. In particular, it includes punctured codes: if $A_n = \text{col}([2^1], \dots, [2^{p-1}])$, then
$\ua_n = (x_{n,1},\dots,x_{n,p-1})$, which is the same as puncturing the first bit, $x_{n,0}$, from the binary image of
$X_n$. Moreover, taking some $t_n = 0$ is equivalent to puncturing the whole symbol $X_n$. The optimization of
puncturing distributions for NB-LDPC codes has been addressed in \cite{ISITA2010}. In this paper, we restrict ourselves
to the case when matrices $A_n$ are of the form $A_n = [I_p \mid B_n ]$, where $I_p$ is the $p\times p$ identity
matrix, meaning that each $\ua_n$ contains the binary image $\ux_n$ (the use of ``extension'' complies with its literal
meaning). We will further assume that any two columns of a matrix $A_n$ are different. It follows that $p\leq t_n \leq
q-1$, and each $\ua_n$ is constituted of the binary image $\ux_n$ and $t_n-p$ pairwise different (nontrivial) extended
bits. In this case, we shall say that {\em the symbol $X_n$ is extended by $k=t_n-p$ bits}.

For instance, let $p=3$ and consider that the following matrix $A$ is used to extend by $2$ bits some coded symbol $X$:
$$
\ua = \ux \times A, \text{ where } A = \left[
\begin{array}{lllll}
    1 & 0 & 0 & 1 & 0 \\
    0 & 1 & 0 & 0 & 1 \\
    0 & 0 & 1 & 1 & 1
\end{array}
\right]
$$
Then $\ua = (\alpha_1, \alpha_2, \alpha_4, \alpha_5, \alpha_6) = (x_0, x_1, x_2, x_0 \wedge x_2, x_1 \wedge x_2)$,
where $\wedge$ denotes the bit-xor operator.

In order to determine the coding rate of the extended code, we denote by $\fdk$ the {\em fraction of degree-$d$ symbols
with $k$ nontrivial extended bits}, $0 \leq k < q - p$; thus
  $\sum_{k=0}^{q-p-1}\!\fdk \!= \!1$.\break
The average number of nontrivial extended bits per coded symbol is given by
  ${f} = \sum_{d=1}^{d_s} \Lambda_d\sum_{k=0}^{q-p-1}k\fdk$,
where $d_s$ is the maximum symbol node degree, and $\Lambda_d = \frac{\lambda_d}{d \int_0^1{\lambda(x)dx}}$ is the
fraction of degree-$d$ symbol nodes. It follows that the extended coding rate is given by $r_e = r\frac{p}{p+f}$. Thus,
we can achieve an arbitrary extended rate within the interval $\left[r \frac{p}{q - 1} , r\right]$ by varying the
parameter $f$.

\begin{figure}[!t]
\centering
\includegraphics[width=\columnwidth]{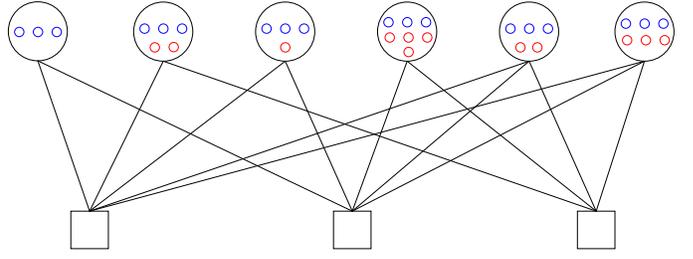}%
\caption{Extension of a NB-LDPC code. Blue circles represent bits of the binary image ($p=3$), while red circles
represent (nontrivial) extended bits.}
\label{fig:extended_codes}%
\vspace{-3mm}
\end{figure}

Figure \ref{fig:extended_codes} illustrates an extended code defined over $\f_8$, with $f_{2,0} = f_{2,1} = f_{2,2} =
f_{2,4} = 1/4$,  and $f_{3,2} = f_{3,3} = 1/2$, which correspond to $f=2$. The mother coding rate is $r=0.5$ and the
extended coding rate is $r_e = 0.3$.

\subsection{Iterative erasure decoding}

We consider that the extended codeword $(\ua_1,\dots,\ua_N)$ is transmitted over a binary erasure channel (BEC). At the
receiver part, the received bits (both from the binary image and extended bits) are used to reconstruct the
corresponding non-binary symbols. Precisely, for each received bit we know its position within the extended binary
image of the corresponding symbol. Hence, for each symbol node we can determine a set of {\em eligible symbols} that is
constituted of symbols whose extended binary images match the received bits. These sets are then iteratively updated,
according to the linear constraints between symbol-nodes \cite{Valentin08}. Alternatively (and equivalently), the
extended code can be decoded by using the linear-time erasure decoding proposed in \cite{Valentin10}.

The {\em asymptotic threshold} of an ensemble of codes is defined as the maximum erasure probability $\pth$ that allows
transmission with an arbitrary small error probability when the code length tends to infinity \cite{Rich-Urba}. Given
an ensemble of codes, its threshold value can be efficiently computed by tracking the fraction of erased messages
passed during the belief propagation decoding; this method is referred to as {\em density evolution}. In this paper,
the density evolution is approximated based on the Monte-Carlo simulation of an infinite code, similar to the method
presented in \cite{ISITA2010}. This method has two main advantages: it can easily incorporate the extending
distribution $\{\fdk\}_{d,k}$, and it can be extrapolated to more general channel models.

% ===================================================================================
  \section{Analysis and optimization}\label{sec:analysis_optimization}
% ===================================================================================

\begin{figure}[!b]
\centering\vspace{-3mm}
\includegraphics[width=\columnwidth]{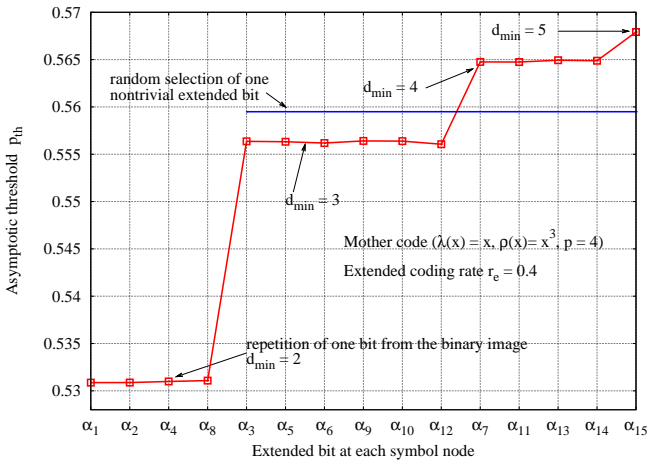}%
\caption{1-bit extension for regular NB-LDPC codes over $\f_{16}$}%
\label{fig:ext_bit_selec_expl}%

\end{figure}

The goal of this section is to answer the following questions.

First of all, assume that we have given a symbol node that has to be extended by $k$ bits. How should they be chosen
among the $q-p-1$ (nontrivial) extended bits?

Secondly, given an extended coding rate $r_e$, how should be extended bits distributed over the symbol-nodes? Put
differently, which is the optimal extending distribution $\left\{\fdk\right\}$?

% ----------------------------------------------------------------------------
  \subsection{Extended bits selection strategy} \label{subsubsec:ext_bit_selec}
% ----------------------------------------------------------------------------
We assume that we have given a symbol-node that has to be extended by $k$ bits. A choice of the $k$ bits among the
$q-p-1$ extended bits corresponds to an extending matrix $A = [I_p \mid B]$ of size $p\times (p+k)$, with pairwise
distinct columns. For each such a matrix, assume that the extended symbol $\ua = \ux \times A$ is transmitted over the
BEC, and let $E(A)$ be the expected number of eligible symbols at the receiver. Recall that an eligible symbol is a
symbol whose extended binary image match the received bits. If all transmitted bits have been erased, any symbol is
eligible. Conversely, if the received bits completely determine the non-binary symbol, then there is only one eligible
symbol. More generally, let $\ua_{\text{rec}}$ denote the sequence of received bits, and $A_{\text{rec}}$ denote the
submatrix of $A$ determined by the columns that correspond to the received positions of $\ua$. Then the eligible
symbols are the solutions of the linear system $\ux \times A_{\text{rec}} = \ua_{\text{rec}}$, and their number is
equal to $2^{p-\text{rank}(A_{\text{rec}})}$. Now, if $\epsilon$ denotes the erasure probability of the BEC, it can be
easily verified that:
  $$E(A) = \sum_{i=0}^{p+k}(1-\epsilon)^i\epsilon^{p+k-i}\left(\sum_{A_i\subseteq A} 2^{p-\text{rank}(A_i)}\right),$$
where the second sum takes over all the submatrices $A_i$ cons\-ti\-tuted of $i$ among the $p+k$ columns of $A$. Hence,
in order to minimize the expected number of eligible symbols $E(A)$, we choose $A$ such that $\dmin(A)$ is maximal,
where $\dmin(A)$ is the smallest number of linearly dependent columns of $A$.

\subsubsection{One extended bit per symbol node} \label{expl:selec_one_ext_bit} Consider the ensemble of regular
$\left(\lambda(x) = x, \rho(x) = x^3\right)$ LDPC codes defined over the $\f_{16}$. Assume that each symbol-node is
extended by $k=1$ bit, such as to achieve an extended rate $r_e = 0.4$. According to the choice of the extended bit
(among the $11$ nontrivial extended bits), $\dmin$ may be equal to $3,4$, or $5$. The asymptotic threshold
corresponding to each choice of the extended bit is shown in Figure \ref{fig:ext_bit_selec_expl}. Note that extended
bits are ordered on the abscissa according to the corresponding $\dmin$. For comparison purposes, we show also the
asymptotic threshold corresponding to the repetition of some bit from the binary image (trivial extended bit
$\alpha_{2^i} = x_i$), in which case $\dmin=2$. Also, the blue line correspond to the threshold obtained if each symbol
node was extended by choosing a random nontrivial extended bit. We observe that the best threshold is obtained when
each symbol node is extended by $\alpha_{15}$, which is\break the XOR of the four bits $x_{0},x_{1},x_{2},x_{3}$ of the
binary image.

\begin{figure}[!b]
\centering \vspace{-3mm}
\includegraphics[width=\columnwidth]{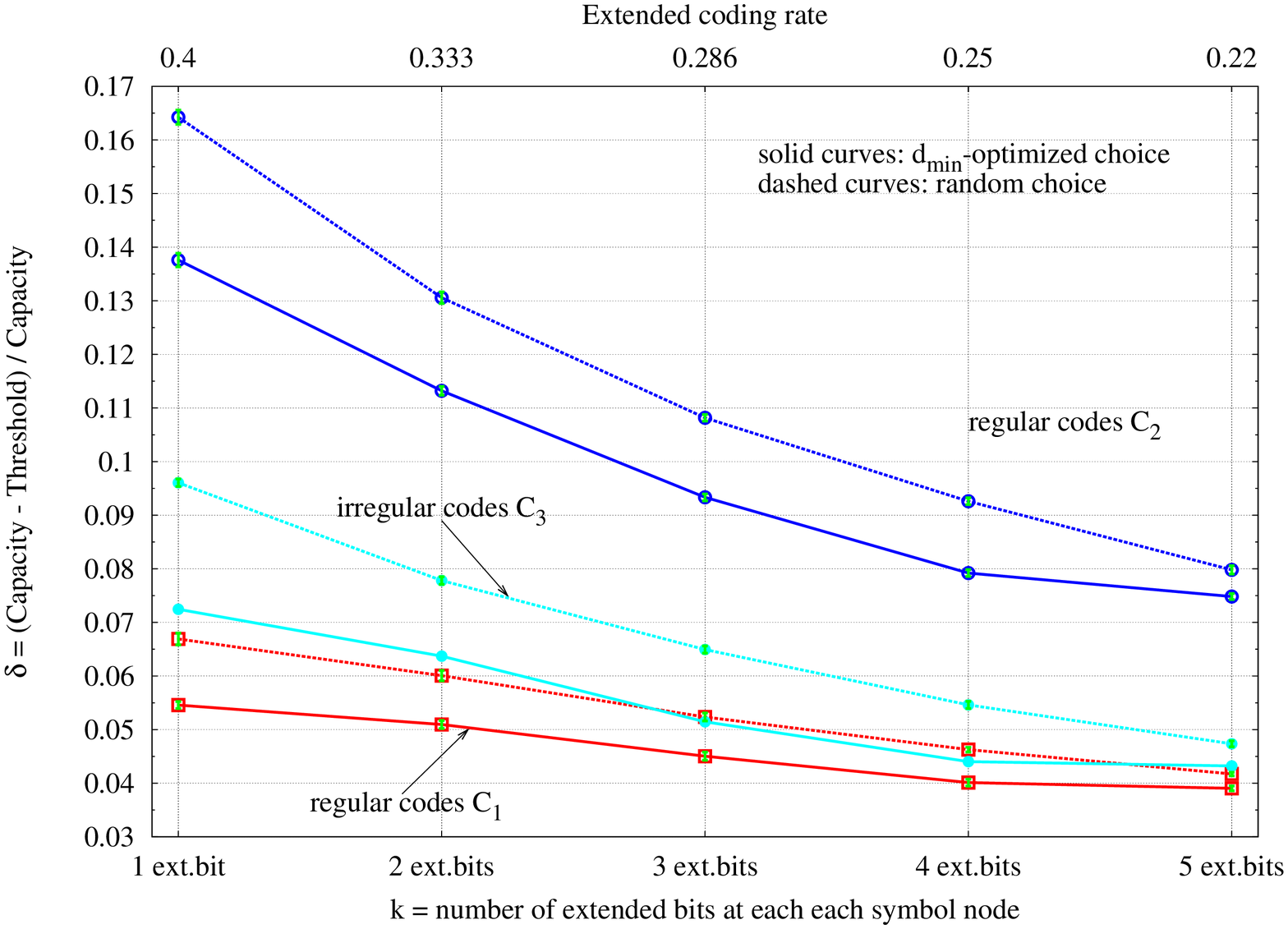}%
\caption{$k$-bits extension for regular and semi-regular NB-LDPC codes over $\f_{16}$}%
\label{fig:ext_bit_selec_expl_overall}%
\end{figure}

\subsubsection{Several extended bits per symbol node} \label{expl:selec_ext_bit}
We consider two ensembles of regular codes $\mathcal{C}_1\left(\lambda(x)=x, \rho(x)=x^3\right)$,
$\mathcal{C}_2\left(\lambda(x)=x^2, \rho(x)=x^5\right)$ and one ensemble of semi-regular codes
$\mathcal{C}_3\left(\lambda(x)=0.5x+0.5x^4, \rho(x)=0.25x^4+0.75x^5\right)$, of coding rate $r=1/2$, defined over
$\f_{16}$. For each ensemble of codes, we consider five different cases, in which all symbol nodes are extended by the
same number $k$ of bits, with $k = 1$, $2$, $3$, $4$,  and  $5$. Accordingly, the extended coding rate $r_e = 0.4$,
$0.33$, $0.29$, $0.25$, and $0.22$.

The {\em normalized gap to capacity}, defined as:
 $$\delta = \frac{\text{capacity} - \text{threshold}}{\text{capacity}} = \frac{1-r-\pth}{1-r},$$
is shown in Figure \ref{fig:ext_bit_selec_expl_overall}. Solid curves correspond to a $\dmin$-optimized choice of the
extended bits, while dashed curves correspond to a random choice of the extended bits. For $k = 5$, there is only a
small difference between these two strategies. However, when $k \leq 4$, the gain of the $\dmin$-optimized choice is
significant for both regular and semi-regular codes.
%Therefore, the $\dmin$-optimized selection of extended bits can improve the
%asymptotic performance of both regular and irregular codes.

% --------------------------------------------------------------------------------
\subsection{Extending distribution analysis} \label{sec:extending_distr}
\begin{figure}[!t]
\centering
\includegraphics[width=\columnwidth]{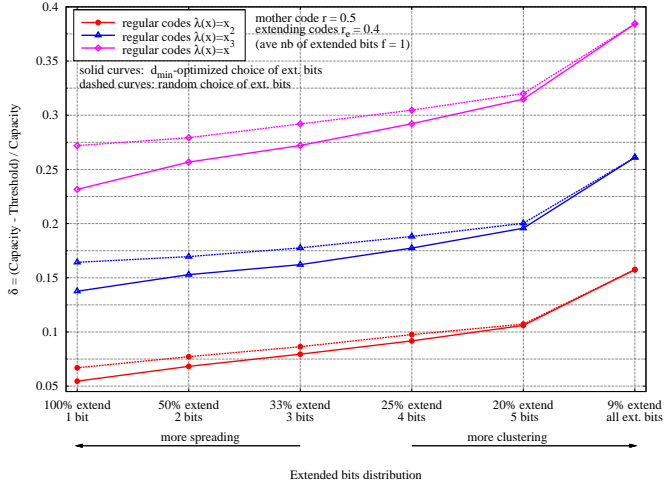}%
\caption{Comparison of spreading vs. clustering extending distributions for regular NB-LDPC codes over $\f_{16}$}%
\label{fig:spread_cluster_regular_expl}%
\vspace{-4mm}
\end{figure}

% --------------------------------------------------------------------------------
First of all, we discuss the case of regular codes. In Figure~\ref{fig:spread_cluster_regular_expl}, we consider three
ensembles of regular codes over $\f_{16}$, with coding rate $r = 0.5$. For each ensemble of codes, we consider five
cases, corresponding to values of $k$ between $1$ and $5$. In each case, a fraction $f_k$ of symbol-nodes are extended
by $k$ bits, while the remaining symbol-nodes have no extended bit. The fraction $f_k$ is chosen such that the extended
coding rate $r_e = 0.4$. Hence, $f_k = 1, 0.5, 0.33, 0.25,0.20,0.09$, for $k = 1,2,3,4,5$, respectively. The right most
point on the abscissa corresponds to a sixth case, in which the extended rate $r_e = 0.4$ is achieved by extending
$9\%$ of symbol-nodes by $k=11$ bits (hence, $k = q-p-1$, which is the maximum number of extended bits). For any of the
three ensembles, we can observe that the smallest gap to capacity is obtained for $k=1$, which means that extended bits
are {\em spread} over as many symbol nodes as possible (in this case, $100$\%), instead of being {\em clustered} over
the smallest possible number of symbol-nodes.

In case of irregular NB-LDPC codes, let $\phi = \left\{f_{d,k}\right\}_{d,k}$ be an extending distribution. Thus,
$\fdk$ is the fraction of degree-$d$ symbols with $k$ nontrivial extended bits, $0 \leq k < (q - p)$. Let ${f}_d$
denote the average number of extended bits per symbol-node of degree-$d$; that is:
 $$ {f}_d = \sum_{k = 0}^{q-p-1}k\fdk \in [0, q-p-1]$$

\noindent We say that {\em the extending distribution $\phi$ is of spreading-type} if for any degree $d$, $f_{d,k} \neq
0$ only if $k=\lfloor{f}_d\rfloor$ or $k=\lceil{f}_d\rceil$. In different words, for any degree $d$, the extended bits
are uniformly spread over all the symbol-nodes of degree $d$. Clearly, a spreading-type distribution is completely
determined by the parameters $\{{f}_d\}$, as we have $f_{d,\lfloor{f}_d\rfloor} = \lceil{f}_d\rceil - {f}_d$,
$f_{d,\lceil{f}_d\rceil} = \lfloor{f}_d\rfloor - {f}_d$, and $f_{d,k} = 0$ for $k \not\in \{\lfloor{f}_d\rfloor,
\lceil{f}_d\rceil\}$.

\smallskip
\noindent We say that {\em the extending distribution $\phi$ is of clustering-type} if for any degree $d$, $f_{d,k}
\neq 0$ only if $k=q-p+1$. In different words, for any degree $d$, the extended bits are clustered over the smallest
possible fraction of symbol-nodes of degree $d$. Clearly, a clustering-type distribution is completely determined by
the parameters $\{{f}_d\}$, as we have $f_{d,q-p+1} = \frac{{f}_d}{q-p+1}$ and $f_{d,k} = 0$ for $k \neq q-p+1$.

\begin{figure}[!t]
\centering
\includegraphics[width=\columnwidth]{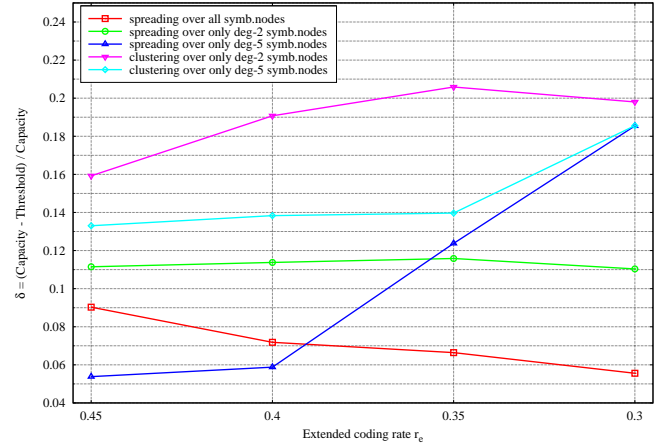}%
\caption{Comparison of several extending-distributions, for semi-regular NB-LDPC codes over $\f_{16}$}%
\label{fig:ext_distr_irreg_expl}%
\vspace{-4mm}
\end{figure}

Now, let us consider the ensemble of semi-regular LDPC codes over $\f_{16}$ with edge-perspective degree distribution
polynomials $\lambda(x) =0.5x+0.5x^4$ and $\rho(x) =0.25x^4+0.75x^5$. The mother coding rate is $r=0.5$, and we intend
to extend symbol-nodes such as to achieve extended coding rates $r_e \in \{0.45, 0.4, 0.35, 0.3\}$. Several extending
distributions are compared in Figure \ref{fig:ext_distr_irreg_expl}. There are three spreading-type distributions,
which spread the extended bits over all the symbol-nodes, or only over the symbol-nodes of degree either $2$ or $5$,
and two clustering-type distributions, which cluster the extended bits over the symbol-nodes of degree either $2$ or
$5$. In all cases, extended bits (or, equivalently, extending matrices $A_n$) are chosen such as to maximize the
corresponding $\dmin$ values. We observe that the smallest gap to capacity is obtained for extending distributions that
spread extended bits either over the degree-$5$ symbol nodes only ($r_e = 0.45,0.4$), or over all the symbol-nodes
($r_e = 0.35,0.3$).
%Therefore, we can state that the spreading distribution is better than clustering distribution for irregular codes.
%However, the spreading of extended codes has to take into account the degree of symbol nodes.

% --------------------------------------------------------------------------------
\subsection{Extending distribution optimization} \label{subsec:strategy}
% --------------------------------------------------------------------------------
Based on the above analysis, we only consider spreading-type extending distributions. Such an extending distribution is
completely determined by the parameters $\{{f}_d\}$, and the extended coding rate can be computed by $r_e = \frac{r}{1+
\frac{1}{p}\sum_{d} \Lambda_d {f}_d}$, where $\Lambda_d$ is the fraction of symbol-nodes of degree $d$.

For given degree distribution polynomials $\lambda$ and $\rho$, and a given extending rate $r_e$, we use the
differential evolution algorithm \cite{DiffEvol} to search for parameters $\{{f}_d\}$ that minimize the asymptotic gap
to capacity. We assume that, for each symbol-node, the extended bits are chosen such as to maximize the corresponding
$\dmin$. The optimized extended codes are presented in the next section.

% ========================================================================
  \section{Code Design and Performance}\label{sec:code_design_performance}
% ========================================================================
In this section we present optimized extending distributions for an irregular mother code over $\f_{16}$. The mother
code has coding rate $r = 1/2$, and it has been optimized by density evolution. The asymptotic threshold is $\pth =
0.4945$, and the edge-perspective degree distribution polynomials are:
$$ \begin{array}{l}
    \lambda(x) = 0.596 x + 0.186 x^4 + 0.071 x^7 + 0.147x^{17} \\
    \rho(x) = 0.2836 x^4 + 0.7164 x^5
\end{array}$$
We optimized extending distributions for extended rates $r_e\in\{0.45, 0.40, 0.35, 0.30, 0.25, 0.20\}$. Optimized
distributions $\{{f}_{d}\}$ are shown in Table \ref{tab:opt_ext_distr}, together with the corresponding asymptotic
threshold $\pth$ and normalized gap to capacity $\delta$. For comparison purposes, we have also indicated the
normalized gap to capacity $\delta_{\text{rand}}$ corresponding to a random choice of extended bits. The last column
corresponds to extended rate $r_e = 2/15$, obtained by extending each symbol-node by the maximum number of extended
bits, {\em i.e.} $q-p-1 = 11$ bits. It can be observed that the optimized distributions allow to maintain an almost
constant value of $\delta\approx 0.01$, for all extended rates $0.45 \geq r_e \geq 2/15$.

\begin{table}[!t]
\centering \caption{Optimized extending distributions for a mother NB-LDPC code with $r=0.5$ over $\f_{16}$, for $r_e =
\left\{0.45\right.$, 0.4, 0.35, 0.3, 0.25, $\left.0.2\right\}$} \label{tab:opt_ext_distr}%
\begin{tabular}{|@{\ }l@{\ }|*{8}{@{\ }c@{\ }|}}
\hline
\textbf{$r_e$} & \textbf{0.5} & \textbf{0.45} & \textbf{0.4} & \textbf{0.35} & \textbf{0.3} & \textbf{0.25} & \textbf{0.2} & \textbf{2/15} \\
\hline
\hline
$f_2$ & 0 & 0.4610 & 1.0164 & 1.7851 & 2.7442 & 4.1290 & 6.1737 & 11 \\
\hline
$f_5$ & 0 & 0.3731  & 1.2113 & 1.2981 & 2.5055 & 3.5864 & 5.3409 & 11 \\
\hline
$f_8$ & 0 & 0.2487  & 0.0359 & 1.8748 & 1.6831 & 2.3393 & 4.7494 & 11 \\
\hline
$f_{18}$ & 0 & 0.1309 & 0.4871 & 0.8511 & 1.6415 & 2.9800 & 4.0234 & 11 \\
\hline
\hline
$\pth$ & 0.4945 & 0.544 & 0.5939 & 0.6406 & 0.69 & 0.74 & 0.7872 & 0.8543 \\
\hline
$\delta$ & 0.011 & 0.0109 & 0.0102 & 0.0145 & 0.0143 & 0.0133 & 0.016 & 0.0143 \\
\hline
\hline
$\delta_{\text{rand}}$ & 0.011 & 0.0234 & 0.0284 & 0.0266 & 0.0251  &  0.0213 & 0.0172  & 0.0143 \\
\hline
\end{tabular}
\vspace{-2mm}
\end{table}

\begin{figure}[!t]
\centering
\includegraphics[width=\columnwidth]{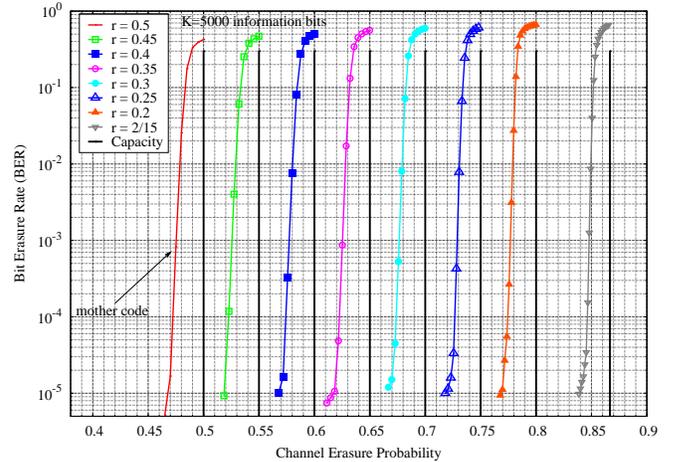}%
\vspace{-2mm}
\caption{Finite length performance of optimized extended NB-LDPC codes}%
\label{fig:flp_opt_ext_distr_irreg}%
\vspace{-3mm}
\end{figure}

Finally, Figure \ref{fig:flp_opt_ext_distr_irreg} presents the Bit Erasure Rate (BER) performance of optimized
extending distributions for finite code lengths. All the codes have binary dimension (number of source bits) equal to
5000 bits (1250 $\f_{16}$-symbols). The mother code with rate $1/2$ has been constructed by using the Progressive Edge
Growth (PEG) algorithm \cite{PEG}, and symbol nodes have been extended according to the optimized distributions
(extension matrices $A_n$ being chosen such as to maximize $\dmin(A_n)$).

% ============================================
  \section{Conclusions}\label{sec:conclusions}
% ============================================
Based on the extended binary image of NB-LDPC codes, we presented a coding technique that allows to produce extra
redundant bits, such as to decreases the coding rate of a mother code. The proposed method allows for using the same
decoder as for the mother code: extra redundant bits transmitted over the channel are only used to ``improve the
quality of the decoder input''.

Extending distributions for regular and irregular codes have been analyzed by using simulated density evolution
thresholds of extended codes over the BEC. We have also presented optimized extending distributions, which exhibit a
normalized gap to capacity $\delta\approx 0.01$, for extended rates from $0.45$ to $2/15$

Finally, although this paper dealt only with NB-LDPC codes over the BEC, the results presented here can be easily
generalized to different channel models.

% Bibliography
% =============================================================== %
  \bibliographystyle{IEEEbib} % bibliography style %
  \bibliography{}          %
% =============================================================== %

\end{document}